\documentclass[smallextended]{svjour3}  
\RequirePackage{fix-cm}
\smartqed  % flush right qed marks, e.g. at end of proof
\usepackage{graphicx,pstricks}
\usepackage{graphics}
\usepackage{moreverb}
\usepackage{subfigure}
\usepackage{epsfig}
\usepackage{subfigure}
\usepackage{hangcaption}
\usepackage{txfonts}
\usepackage{palatino}
\usepackage{url}
\def\Diag{{\rm Diag}}

\def\trace{{\rm Trace}}

\def\R{\mathbb{R}}

\def\P{\mathcal{P}}
\def\D{\mathcal{D}}
\def\X{\mathcal{X}}
\def\O{\mathcal{O}}

\def\SRn{{\it S}\R^{n \times n}}
\def\beq{\begin{equation}}
\def\eeq{\end{equation}}

\def\eps{\epsilon}

\newtheorem{assumption}{Assumption}
\newtheorem{algorithm}{Algorithm}

\def\ba{\begin{array}}
\def\ea{\end{array}}
\def\beann{\begin{eqnarray*}}
\def\eeann{\end{eqnarray*}}
\def\bea{\begin{eqnarray}}
\def\eea{\end{eqnarray}}

\def\BT{\begin{theorem}}
\def\ET{\end{theorem}}
\def\BL{\begin{lemma}}
\def\EL{\end{lemma}}
\def\BP{\begin{proposition}}
\def\EP{\end{proposition}}
\def\BC{\begin{corollary}}
\def\EC{\end{corollary}}
\def\BD{\begin{definition}}
\def\ED{\end{definition}}
\def\BA{\begin{assumption}}
\def\EA{\end{assumption}}
\def\BR{\begin{remark}}
\def\ER{\end{remark}}
\def\BE{\begin{example}}
\def\EE{\end{example}}
\def\BAL{\begin{algorithm}}
\def\EAL{\end{algorithm}}

\begin{document}

\title{A First-Order Algorithm for the A-Optimal Experimental Design Problem: A Mathematical Programming Approach\footnote{Supported by Start-Up Grant No. SRG ESD 2012 033, SUTD}}
\titlerunning{On the A-Optimal Experimental Design Problem} 
\author{Selin Damla Ahipa\c{s}ao\u{g}lu
}
\institute{S. D. Ahipa\c{s}ao\u{g}lu \at
              Singapore University of Technology and Design       \\
        \email{ahipasaoglu@sutd.edu.sg}           %  \\
}
\maketitle
 \thispagestyle{empty}

%%%%%%%%%%%%%%%%%%%%%%%%%%%%%%%%%%%%%%%%%%%%%%%%%%%%%%%%%%%%%%%%%%%%%%%%%%%%%%%%%%%%%%%%%%%%%%%%%%%%%%%%%%%%%%%%%%%%%%%
%%%%%%%%%%%%%%%%%%%%%%%%%%%%%%%%%%%%%%%%%%%%%%%%%%%%%%%%%%%%%%%%%%%%%%%%%%%%%%%%%%%%%%%%%%%%%%%%%%%%%%%%%%%%%%%%%%%%%%%
%%%%%%%%%%%%%%%%%%%%%%%%%%%%%%%%%%%%%%%%%%%%%%%%%%%%%%%%%%%%%%%%%%%%%%%%%%%%%%%%%%%%%%%%%%%%%%%%%%%%%%%%%%%%%%%%%%%%%%%
\begin{abstract}
We develop and analyse a first-order algorithm for the A-optimal experimental design problem. The problem is first presented as a special case of a parametric family of optimal design problems for which duality results and optimality conditions are given. Then, two first-order (Frank-Wolfe type) algorithms are presented, accompanied by a detailed time-complexity analysis of the algorithms and computational results on various sized problems.
\end{abstract}

%%%%%%%%%%%%%%%%%%%%%%%%%%%%%%%%%%%%%%%%%%%%%%%%%%%%%%%%%%%%%%%%%%%%%%%%%%%%%%%%%%%%%%%%%%%%%%%%%%%%%%%%%%%%%%%%%%%%%%%
%%%%%%%%%%%%%%%%%%%%%%%%%%%%%%%%%%%%%%%%%%%%%%%%%%%%%%%%%%%%%%%%%%%%%%%%%%%%%%%%%%%%%%%%%%%%%%%%%%%%%%%%%%%%%%%%%%%%%%%
%%%%%%%%%%%%%%%%%%%%%%%%%%%%%%%%%%%%%%%%%%%%%%%%%%%%%%%%%%%%%%%%%%%%%%%%%%%%%%%%%%%%%%%%%%%%%%%%%%%%%%%%%%%%%%%%%%%%%%%

\section{Optimal Experimental Design}
\label{intro}

Consider the following linear model
 \bea \label{eq:mod_rand} y = x^T(t) \theta + \eps(t),\eea
where components of $x^T(t) = (x_1(t), x_1(t), \dots, x_n(t))$ are $n$ linearly independent continuous functions on some compact space and $\theta\in \R^n$ is a vector of unknown parameters to be estimated. Let the error terms $\eps(t)$ follow a multivariate normal distribution with mean 0 and the error in each observation be independent from the others. Without loss of generality, suppress the dependency of the vector $x(t)$ on the actual experimental conditions $t$ and work with a model function such as \bea\label{eq:lin_mod} y=x^T\theta +\epsilon,\eea in which the vector $x$ will be referred to as the regression or design vector. Let $\mathcal{X}=\{x_1,\dots, x_m\}\subset \R^n$ be the set of regression vectors, assume henceforth that the $x_i$'s span $\R^n$, and $X$ denote a matrix of order $n\times m$ whose columns consist of these vectors. (Frequently, the regression points are chosen from
some fixed compact set, here suppose
that some large fixed subset $\cal{X}$ has been preselected.) 

\BD An \emph{experimental design of size $N$} is given by a finite number of regression points $x_1,\dots,x_m$ in $\R^n$ and nonnegative integers (representing the number of repetitions at each respective point) $n_1,\dots,n_m$ such that $\sum_{i=1}^m{n_i}=N$.\ED

%If we assume that one experiment is conducted at each of these regression points and the model at hand is a general linear model with the normality assumption discussed in the previous paragraph, it is well-known that the optimal (unbiased) estimator for the parameter vector %$\theta$ is $(X^TX)^{-1}X^TY$, and has dispersion matrix $\sigma^2(X^TX)^{-1}$. It may be desirable to conduct multiple or zero experiments at some of the regression points, and so we have the following definition:

%In an experimental design, $n_i$ represents the number of repetitions of the experiment at the regression point $x_i$.  
In this setting, the \emph{dispersion matrix} related to the optimal (unbiased) estimator for the parameter vector is \bea\label{eq:disp} D=\sigma^2\left(\sum_{i=1}^m{n_ix_ix_i^T}\right)^{-1}=\frac{\sigma^2}{N}\left(\sum_{i=1}^m{\frac{n_i}{N}x_ix_i^T}\right)^{-1}.\eea \emph{Optimal experimental design} focuses on finding integers $n_i$ so that the dispersion matrix, which is a measure of the variance (or the error) of the estimator, is minimized in some sense. The dispersion matrix is positive definite (i.e., $D\succ 0$) and usually the minimization is with respect to the Loewner ordering over the cone of positive semidefinite matrices
($A\succeq B \iff A-B \in \SRn_+$). Since this is an antitonic ordering, minimizing the dispersion matrix is equivalent to maximizing the \emph{information matrix} $$M=\frac{N}{\sigma^2}\sum_{i=1}^m{\frac{n_i}{N}x_ix_i^T}.$$ 

%%It is easy to see that the dispersion matrix $D$ belongs to the cone of symmetric positive semidefinite matrices, $\SRn_+$. It is reasonable to be interested in the matrices which are minimal 

%We also write $\SRn_{++}$ for the cone of symmetric positive definite matrices and write $A \succ B$ to mean $A-B \in \SRn_{++}$. Since the Loewner ordering is antitonic, i.e., $A\succeq B$ implies that $ B^{-1}\succeq A^{-1}$, minimizing $D$ is equivalent to maximizing %the \emph{information matrix} $$M(u):=\frac{N}{\sigma^2}\sum_{i=1}^m{\frac{n_i}{N}x_ix_i^T} $$ 

When the total number of experiments $N$ is finite, experimental design problems become integer programming problems which are quite hard to attack especially for large $m$. Hence the case where $N$ tends to infinity is studied instead. In this case we maximize $M(u):=\sum_{i=1}^m{u_ix_ix_i^T}$, where $u_i \geq0$, for $i = 1,\dots, m$, and  $\sum_{i=1}^m{u_i} = 1$. Note that an experimental design with an infinite sample size $N$ defines a probability distribution which assigns all its weight to a finite number of points. The points with positive weight are the \emph{support points} of the experimental design. One can refer to Chapter 12 in \cite{Puk93} or \cite{TMM09} for a valuable discussion on how to come up with an exact experimental design for a finite sample size once the optimal design for an infinite sample size is found. 

%When solving experimental design problems, finding a solution which is minimal with respect to the Loewner ordering is necessary. Nevertheless, optimization using this criterion is not straightforward and the set of matrices that satisfy it is usually very large. 
%We prefer real-valued criteria which still preserve Loewner optimality. 
\BD An \emph{information function} is a function $\phi$ from the cone of positive semidefinite matrices to the real line,  $\phi: \SRn_+ \rightarrow \R,$ which is positively homogeneous, superadditive, nonnegative, nonconstant, and upper semicontinuous. \ED 
It is easy to see that information functions are concave. They order the information matrices according to their informative value and preserve the Loewner ordering. The most common information functions are matrix means. 
\BD Let $\lambda(C)$ denote the eigenvalues of a matrix $C$. If $C$ is a positive definite matrix, i.e., $C\succ0$, the matrix mean $\phi_p$ is a function defined as
$$\phi_p(C)=\left\{
\ba{ccc}
	\lambda_{\max}(C) & {\rm for } & p=\infty;\\
	\left(\frac{1}{n}\trace C^p\right)^{1/p} & {\rm for }  &p\neq0,\pm\infty;\\  
	(\det C)^{1/n} & {\rm for }  &p=0;\\  
	\lambda_{\min}(C) & {\rm for }  &p=-\infty.
\ea\right.
$$ If $C$ is a singular positive semidefinite matrix, then 
$$\phi_p(C)=\left\{
\ba{ccc}
	\lambda_{\max}(C) & {\rm for } & p=\infty;\\
	\left(\frac{1}{n}\trace C^p\right)^{1/p} & {\rm for }  &p\neq0,\infty;\\  
	0 & {\rm for }  &p\leq0.\\  
\ea\right.
$$
\ED

Matrix means satisfy the necessary properties of information functions when $p\leq1$. Using these functions, the \emph{general optimal experimental design problem} is defined as follows:
$$
\ba{rrrcl}
  & \max_u          & g_p(u)& :=& \ln \phi_p(M(u))\\
  (\mathcal{D}_p)  & &          e^T u & =    & 1, \\
       &             &     u & \geq & 0,
\ea$$ where $e$ is a vector of ones in $\R^m$.
Each value of the parameter $p$ gives rise to a different criterion with different applications. We will study one of the special cases (when $p=-1$) in great detail in Section \ref{sec:Aopt} forward.

\section{Ellipsoidal Inclusion Problems}

Assume that we have a set of points $\mathcal{X}=\{x_1,\dots,x_m\}\subset \R^n$, which spans $\R^n$ and is symmetric with respect to the origin. We are interested in approximating (especially enclosing) the convex hull of these points with an ellipsoid. Note that the idea is to approximate the complex structure of the convex hull with a simple geometric object. Boxes, balls, ellipsoids, and cylinders are used in the literature. Ellipsoids are preferred in many applications since they are smooth and flexible, and testing membership in or optimizing a linear function over an ellipsoid is a straightforward task.

The set $$\mathcal{E}(\bar{x},H):=\{x\in\R^n : (x-\bar{x})^T H (x-\bar{x})\leq n\}$$ for $\bar{x}\in \R^n$ and $H\succ0$ is an ellipsoid in $\R^n$. It is centered at $\bar{x}$ and its shape is defined by $H$. It can be viewed as a unit ball under an affine map where each point $\tilde{x}$ in the unit ball is mapped to a point $x=\bar{x}+\sqrt{n}L\tilde{x}$ in the ellipsoid, where $L$ satisfies $LL^T=H^{-1}$. Geometric properties of the ellipsoid such as its volume, length of its semi-axes, etc., are determined by the shape matrix $H$. For example, its volume is $\frac{n^{n/2}}{\sqrt{\det{H}}}$ times that of the unit ball. 

The convex hull of a set of finitely many points can be enclosed by an infinite number of ellipsoids. Obviously we are only interested in ellipsoids which are centered at the origin (since $\mathcal{X}$ is symmetric around the origin) and resemble the convex hull in some sense. Although the enclosing ellipsoid which has the minimum volume is a natural choice from both theoretical and practical points of view, as discussed in detail in \cite{TYil07} and \cite{AST08}, defining the problem using a more general criterion is quite insightful since other criteria can be needed in certain applications.

For $q\leq 1$, consider the following problem:
$$\ba{rrrcl}
  & \min_H & f_q(H) &:=& - \ln \phi_q(H) \\
  (\mathcal{P}_q)  &             & x_i^T H x_i & \leq & n, \, i =
  1,\dots,m,\\
  & & H & \succ & 0.
\ea $$ For each value of $q$, this problem finds an ellipsoid which encloses all points in $\mathcal{X}$, is centered at the origin, and has a shape matrix with the largest matrix mean $\phi_q$. Each value of the parameter $q$ leads to a different problem with a different geometric interpretation. For example, when $q=0$, the objective function becomes (a multiple of) $\ln \det(H^{-1})$ and hence $(\mathcal{P}_q)$ is equivalent to the Minimum-Volume Enclosing Ellipsoid problem discussed in the previously mentioned references. Similarly, for the extreme case of $q=-\infty$, we have $\ln (\lambda_{\min}(H))^{-1}$ as the objective function and hence the problem becomes that of finding the Minimum Enclosing Ball of $\mathcal{X}$. (See F\cite{Yil08} and \cite{AY08} for efficient algorithms for this problem.) When $q=1/2$, $(\mathcal{P}_q)$ maximizes the trace of $H^{1/2}$ and leads to a less familiar geometric problem in which we would like to maximize the sum of the inverses of the semi-axes of the enclosing ellipsoid. This problem has important applications in statistics and solving this problem is the main topic of this paper. We will refer to the general problem $(\mathcal{P}_q)$ as the \emph{ellipsoidal inclusion} problem.

\section{Duality}

We now show that the two problems introduced above are closely related.
 
\BL\label{lemm:weak}{\rm [Weak Duality]} Let $p$ and $q$ be a pair of conjugate numbers in $\left(-\infty,1\right)$, i.e., they satisfy $pq=p+q$. Then we have $f_q(H)\geq g_p(u)$ for any $H$ and $u$
feasible in $(\mathcal{P}_q)$ and $(\mathcal{D}_p)$, respectively.\EL 
\proof We have
\beann f_q(H)-g_p(u) & = & - \ln\phi_q(H) - \ln\phi_p(M(u)) \\
& = & - \ln \left(\phi_q(H)\phi_p(M(u))\right)\\
&\geq &-\ln\left(\frac{1}{n}H\bullet M(u)\right)\\
&\geq& -\ln 1=0,
\eeann where $\bullet$ denotes the trace product of two symmetric matrices, i.e., $A\bullet B =\trace (AB)$. The first inequality is an application of the H\"{o}lder's inequality (on the eigenvalues of the matrices at hand) and a detailed proof can be found in \cite{Puk93}. The second inequality follows from the feasibility of the solutions $H$ and $u$. Indeed, $\frac{1}{n}H\bullet(M(u)) = \frac{1}{n}\sum_{i=1}^m \left(u_i H\bullet(x_ix_i^T)\right) \leq \frac{1}{n}\sum_{i=1}^m{\left( u_i(x_i^THx_i)\right)}\leq \frac{n}{n}=1.$\qed

\BT \label{theo:opt_cond} {\rm [Strong Duality]} Let $p$ and $q$ be a pair of conjugate numbers in $\left(-\infty,1\right)$. There exist optimal solutions for problems $(\mathcal{P}_q)$ and $(\mathcal{D}_p)$. Furthermore, the following conditions, together with primal and dual feasibility, are necessary and sufficient for optimality in both
$(\mathcal{P}_q)$ and $(\mathcal{D}_p)$: \begin{itemize}
\item[a.] $H=\frac{n}{\trace{(M(u))^{p}}}(M(u))^{p-1}$ and 
\item[b.] $x_i^THx_i=n$ if $u_i>0$.
\end{itemize} \ET
\proof Let $H$ be a feasible solution for problem $(\mathcal{P}_q)$. Summing up
the linear constraints, we must have $\sum_{i=1}^m{x_i^THx_i}=H\bullet
XX^T\leq nm$. Since $XX^T\succ0$ and $nm>0$, $\{H \succeq 0 : H\bullet
XX^T\leq nm\}$ is a compact set. Hence the feasible region for
problem $(\mathcal{P}_q)$ is also a compact set (since it is the intersection of
a compact set with a finite set of halfspaces). Moreover, $H = \epsilon I$
is feasible for $(\P_q)$ for sufficiently small positive $\epsilon$, and we can add the constraint
that $f_p(H) \leq f_p(\epsilon I)$ without loss of generality.
The objective function is (finite and) continuous on this
modified compact feasible region, so an optimal solution
exists for problem $(\mathcal{P}_q)$. Existence of an optimal solution for
$(\mathcal{P}_q)$ implies the existence of an optimal solution for
$(\mathcal{D}_p)$ as will be discussed later.

Sufficiency follows from the previous lemma, since the conditions imply equality in the weak duality inequality. In order to prove necessity, let $\tilde{H}$ be an optimal solution for $(\mathcal{P}_q)$. The KKT conditions must hold for this solution, i.e., there exist nonnegative multipliers $\tilde{u}\in\R^m$ such that the following equalities hold:
\bea \label{eq:hu}-\frac{n}{\trace{\tilde{H}}^{q}}{\tilde{H}}^{q-1}+M(\tilde{u})&=&0,\\\label{eq:comsl}
\tilde{u}_i(n-x_i^T\tilde{H}x_i)&=&0, \quad i=1,\dots,m.
\eea 
These equalities imply that $\sum_{i=1}^m{\tilde{u}_i}=1$, since 
\beann \sum_{i=1}^m{\tilde{u}_i} & = & \frac{\sum_{i=1}^m \tilde{u}_ix_i^T\tilde{H}x_i}{n}\\
& = & \trace \left(\frac{\tilde{H} M(\tilde{u})}{n}\right)\\
& = & \trace \left(\frac{\tilde{H} \left(\frac{n}{\trace{\tilde{H}}^{q}}{\tilde{H}}^{q-1}\right)}{n}\right)\\
& = & \frac{n\trace H^q}{n\trace H^q}=1,\eeann  
and hence $\tilde{u}$ is a feasible solution for $(\mathcal{D}_p)$. Strong duality holds for the solution pair $\tilde{H}$ and $\tilde{u}$, so strong duality holds for any pair of optimal solutions $H$ and $u$. Conditions (a) and (b) are direct consequences of Equations (\ref{eq:hu}) and (\ref{eq:comsl}), and hence they are necessary. \qed

Let $\beta_i(u):=x_i^T(M(u))^{p-1}x_i$. The following identity will be used extensively. \bea\label{eq:base}
u^T\beta(u)&=&\sum_{i=1}^m{u_i\beta_i(u)}\nonumber\\
&=& \sum_{i=1}^m{\trace\left(u_ix_i^T(M(u))^{p-1}x_i\right)}\nonumber\\
&=& \trace\left((M(u))^{p-1}\sum_{i=1}^m{u_ix_ix_i^T}\right)\nonumber\\
&=& \trace{(M(u))^{p}}.
\eea Using (\ref{eq:base}), we can write the necessary and sufficient conditions for $u$ to be optimal in $(\D_q)$ (the optimal
$H$ for $(P_q)$ follows from (a)) as
\begin{itemize}
\item[(i)] $\beta_i(u)\leq {u}^T\beta(u)$ for all $i$, and 
\item[(ii)] $\beta_i(u)={u}^T\beta(u)$ if $u_i>0$,
\end{itemize} which motivates the following definitions.

\BD\label{def:sol} Given a positive $\epsilon$, we call a dual feasible point $u$ an \emph{$\epsilon$-primal feasible solution} if $\beta_i(u)\leq u^T\beta(u)(1+\epsilon)$ for all $i$, and say that it  satisfies the \emph{$\epsilon$-approximate optimality conditions} or it is an \emph{$\epsilon$-approximate optimal solution} if moreover $\beta_i(u)\geq u^T\beta(u)(1-\epsilon)$ whenever $u_i>0$.
\ED

The following lemma justifies the notation and proves that an $\epsilon$-primal feasible solution for $(\mathcal{D}_p)$ is close to being optimal in a well-defined way. 

\BL \label{lem:dual_gap} Let $p$ and $q$ be a pair of conjugate numbers in $\left(-\infty,1\right)$. Given a dual feasible solution $u$ which is $\epsilon$-primal feasible, $H=\frac{n}{(1+\epsilon)\trace{(M(u))^{p}}}(M(u))^{p-1}$ is feasible in $(\mathcal{P}_q)$ and we have 
$0\leq g_p^*-g_p(u)\leq \ln
 (1+\epsilon)$ where $g_p^*$ is the optimal objective function value
 of $(\mathcal{D}_p)$. \EL
 \proof  The $\epsilon$-primal feasibility implies that
$H=\frac{n}{(1+\epsilon)\trace{(M(u))^{p}}}(M(u))^{p-1}$ is feasible for the primal
problem $(\mathcal{P}_q)$.  Let us first assume that $p,q\neq 0$. Then by weak duality, we have 
\beann 0 & \leq & f_q(H)-g_p^* \\
&=&-\frac{1}{q}\ln\left(\frac{1}{n}\trace\left(\frac{n(M(u))^{p-1}}{(1+\epsilon)\trace{(M(u))^{p}}}\right)^q\right)-g_p^*\\
&=&\ln(1+\eps)-\frac{1}{q}\ln\left(\frac {n^{q-1}\trace(M(u))^{(p-1)q}}{(\trace{(M(u))^{p}})^q}\right)-g_p^*\\
&=&\ln(1+\eps)+\ln\left(n^{\frac{1-q}{q}}\left(\trace(M(u))^{p}\right)^{\frac{q-1}{q}}\right)-g_p^*\\
&\leq&\ln(1+\eps)+\frac{1}{p}\ln\left(\frac{1}{n}\trace(M(u))^{p}\right)-g_p^*\\
g_p^*-g_p(u)&\leq&\ln(1+\eps).
\eeann The case where $p=q=0$ is similar and the proof can be found in \cite{AST08}.\qed

\BL \label{lem:init_gapKHA} $u^0=\frac{1}{m}(1,1,\dots,1)$ is an ($m$-$1$)-primal feasible solution.\EL
\proof We have \beann \sum_{i=1}^m\frac{1}{m}\beta_i(u^0)&=& (u^0)^T\beta(u^0), {\rm or}\\
\sum_{i=1}^m\beta_i(u^{0}) & = & m(u^0)^T\beta(u^0), {\rm so \ that}\\
\max_{1\leq i \leq m}\beta_i(u^0) & \leq & (1+(m-1))(u^0)^T\beta(u^0),\eeann and the result follows from the definition of an ($m$-1)-primal feasible solution.\qed

So far, we have developed the duality relation between problems $(\mathcal{P}_q)$ and $(\mathcal{D}_p)$ and characterized the optimal solutions of these problems. We also have an initial solution for $(\mathcal{D}_p)$ which is somewhat close to optimality and we can assess the quality of the solutions at hand. (Note that we will refer to this initialization method as ``Khachiyan's Initialization" since it was used by Khachiyan in \cite{Kha96} for $p=q=0$.) In other words, we know how to start and end an algorithm for $(\mathcal{D}_p)$ and now we need to figure out how to move from a given solution to a better one. The selection of the iterate and the analysis of the algorithm changes with respect to the specific parameter, namely $p$, of the optimal experimental design. In the following section, we will develop a Frank-Wolfe type first-order algorithm for the case when $p=-1$ (and hence $q=1/2$). This problem is referred to as the A-optimal experimental design in statistics.

\BR \label{rem:lit} We would like to note that most of the results in this section are not entirely new to the statistic community. What is new, and hopefully useful, is the treatment of the subject using a standard mathematical programming approach that builds the necessary machinery in devising algorithms and analysing their convergence  properties. %The main goal of this paper is illustrating the use of this approach for a specific pair of parameters $p$ and $q$, and obtain new theoretical results. 
Specifically, (i) Theorem \ref{theo:opt_cond} in this section (and Theorem \ref{theo:opt_condTR} below, which is a special case of Theorem \ref{theo:opt_cond}) can be obtained by following Theorems 7.12, 7.19, and 7.20 in \cite{Puk06}; and (ii) Lemma \ref{lem:dual_gap} is similar to Proposition IV.28 in \cite{Paz86}. Instead of borrowing these results directly from literature, we have provided a consistent and comprehensive treatment of the subject here. We strongly believe this is a simpler and -in some sense- more intuitive approach for building algorithms. Understanding the relationship between primal and dual problems, and the derivation of the optimality conditions based on this relation is necessary to follow the rest of the paper. One exception is possibly Lemma \ref{lem:init_gapKHA}, which was only obtained for the $p=q=0$ case in \cite{Kha96}. The generalized result provided here is novel according to our knowledge.
\ER

\BR \label{rem:geo} The duality relationship between problems  $(\mathcal{P}_q)$ and  $(\mathcal{D}_p)$ presented in this chapter, provides a geometric and non-trivial insight to the design problem: Finding the best experimental design is equivalent to covering the induced design space with a 'minimum volume' ellipsoid (where the measure of the volume is dictated by the criterion used for the design problem). This interpretation is also well-known to the statistics community for the case $p=q=0$. (See: \cite{S72},\cite{ST73}, \cite{H93}, and more recently in \cite{AST08}). In \cite{BDZ06}, the authors provide a similar discussion about the geometric interpretation of $ (\mathcal{D}_p)$-optimal design problems for all values of $p$ for models with two parameters. Our discussion is more general since it is independent of the number of parameters in the model. Understanding the geometric interpretation plays a significant role in internalizing several pieces of the machinery developed in this paper, especially in construction of approximate solutions, quantification of the duality gap associated with them and choosing pivots for the algorithm. A similar geometric interpretation exists for the ${\rm D}_k$-optimal experimental design problem: A generalization of the D-optimal experimental design problem where we are only interested in estimating the first $k$ out of $n$ parameters in a general linear model. In this case, finding the best experimental design is equivalent to covering the induced design space with a minimum-area ellipsoidal cylinder with special properties about its base and axis as discussed in \cite{AT13}. It is easy to see that although this paper discusses only the D-criterion, the geometric interpretation carries to other criteria in a straightforward way. 
\ER

\BR Finally, before continuing our discussion towards algorithms for the A-optimal experimental design problem ($p=-1$ and $q=1/2$) below, we would like to mention that `in principle' algorithms for problems with other values of $p$ (and respective $q$) can be designed and analysed following the steps outlined here. Nevertheless, the step sizes and convergence analysis need to be customized for each criterion, and can be challenging in some cases. One can refer to \cite{AST08} for a detailed analysis of similar algorithms for the D-optimal experimental design problem. \ER

%%%%%%%%%%%%%%%%%%%%%%%%%%%%%%%%%%%%%%%%%%%%%%%%%

%%%%%%%%%%%%%
\section{Existing Algorithms} \label{sec:LIT}
%%%%%%%%%%%%%

Many Frank-Wolfe type algorithms have been devised to solve experimental design problems, especially for the D-optimal experimental design problem. Some of these were developed by statisticians: \cite{Fed72} and \cite{W72} provided algorithms that maximize a linearization of the objective function over the unit simplex at each iteration. These algorithms only allow iterations that increase the weight of one of the coordinates of the solution. These were improved significantly by \cite{Atw73} where decreasing the weight of the chosen coordinate was also considered, paralleling the addition of \emph{Wolfe's Away Steps} to Frank's algorithm (see \cite{FW56} for the original Frank-Wolfe algorithm). Recently, these algorithms were analysed rigorously by the optimization community, motivated by the ellipsoidal inclusion problem rather than the design problem. The algorithms in \cite{Kha96} and \cite{KumYil05} were equivalent to that of \cite{Fed72}. In addition, \cite{KumYil05} proposed an initialization scheme that produces optimal solutions with significantly smaller number of nonzero weights than previous algorithms. This was accompanied by introducing the concept of \emph{core sets}, and the authors were able to provide upper bounds on the number of nonzero weights in the optimal design. Later, \cite{TYil07} extended the analysis to include Wolfe's away steps, hence providing rigorous complexity results for an algorithm equivalent to that of \cite{Atw73}. During this period, \cite{HarPro07} proved a simple condition that can be used to identify and eliminate points that do not lie on the boundary of the optimal ellipsoid, i.e., points that are guaranteed to have zero weight in the optimal design. (Recently, this result has been extended for all values of $p$ in \cite{HarPro13}.) Incorporating this condition to any Frank-Wolfe type algorithm is very easy and improves the computational time significantly (see Chapter 2 in \cite{Ahi09}). In addition, \cite{AST08} proved that the Frank-Wolfe type algorithms with an exact line search have favorable local convergence properties and therefore can be used to obtain very accurate solutions. In the following section, we will devise and analyse an algorithm which is a Frank-Wolfe type algorithm with Wolfe's aways steps. It can be viewed as applying Atwood's approach to the A-optimal experimental design problem. The global and local convergence properties that will be established below are in line with those developed recently by the optimization community for the D-optimal experimental design problem.

In contrast to Frank-Wolfe type algorithms, multiplicative algorithms update all weights simultaneously. Several versions were developed for various criteria: C-optimality in \cite{Fell74}, D-optimality in \cite{Tit76}, and A-optimality in \cite{Tor83}. Recently, faster algorithms were developed in \cite{Yu11} for D-optimality and in \cite{YBT13} for the general experimental design problem, i.e., problem ($\mathcal{D}_p$) discussed in this paper. A relatively recent survey on multiplicative algorithms together with a new multiplicative approach can also be found in \cite{TMM09}.

Another interesting and modern approach to the experimental design problem is using semidefinite programming reformulations as discussed in \cite{VB99}. This approach fails to solve large problems due to the lack of efficient solvers as demonstrated in Section \ref{sec:TRcompSDP}.

%%%%%%%%%%%%%%%%%%%%%%%%%%%%%%%%%%%%%%%%%%

\section{The A-Optimal Experimental Design Problem}\label{sec:Aopt}
Let $\mathcal{X}=\{x_1,\dots, x_m\}\subset \R^n$ be a set of regression vectors and $X$ denote a matrix of order $n\times m$ whose columns consist of these vectors. Finding a design which minimizes the mean dispersion of the parameters in (\ref{eq:mod_rand}) amounts to solving $$
\ba{rrrcl}
  & \max_u           & \hat{g}(u) &:=& -\trace (M(u))^{-1} \\
  (\hat{\mathcal{D}})  &             & e^T u & =    & 1, \\
       &             &     u & \geq & 0,
\ea
$$
where $e$ is a vector
of ones in $\R^m$ as in the previous sections. Problem $(\hat{\mathcal{D}})$ is referred to as the A-optimal experimental design problem in statistics. In \cite{Fed72}, Fedorov proved that a Frank-Wolfe type algorithm converges to an optimal design and discussed the conditions under which D-optimal and A-optimal designs coincide. In this paper, we will introduce a pair of problems dual to each other and closely related to $(\hat{\D})$. Using the interplay between these problems, we will develop various Frank-Wolfe type algorithms and prove that an $\eps$-approximate solution (defined as in Section \ref{intro}) can be obtained in $\mathcal{O}(n\ln n+\epsilon^{-1})$ or $\mathcal{O}(\ln m+\epsilon^{-1})$ iterations. Each step of the algorithm can be performed in $\mathcal{O}(nm)$ arithmetic operations. In Section \ref{sec:TRloc}, we will prove that some of these algorithms possess a local linear convergence property. These algorithms are also preferable in practice as illustrated by the computational results in Section \ref{sec:TRcomp}.

Consider the following two problems:
$$\ba{cccl}
   \min & f(H) := - 2\ln \trace H^{1/2} \\
  (\mathcal{P})  &             x_i^T H x_i \leq  1, \, i =
  1,\dots,m,
\ea$$ and
$$
\ba{cccl}
   \max_{u}          & g(u) := -\ln \trace (M(u))^{-1}  \\
  (\mathcal{D})  &             e^T u  =     1, \\
         &                 u  \geq  0.
\ea
$$ $(\mathcal{P})$ is a special case of $(\mathcal{P}_q)$ in Section \ref{intro} in which $q=1/2$. From a geometric point of view, it is the problem of finding an ellipsoid which encloses all data points in $\X$ and has the largest sum of inverses of its semi-axes. Also $(\mathcal{D})$ is a special case of $(\mathcal{D}_p)$ in Section \ref{intro} where $p=-1$. This problem is equivalent to the statistical problem $(\hat{\mathcal{D}})$ introduced above. We will use both $(\mathcal{D})$ and $(\hat{\mathcal{D}})$ in order to develop and analyze first-order algorithms for solving all of the three problems mentioned above simultaneously. We will first establish weak duality:
\BL\label{lemm:weakTR}{\rm [Weak Duality]} We have $f(H)\geq g(u)$ for any $H$ and $u$ feasible in $(\mathcal{P})$ and $(\mathcal{D})$, respectively.\EL
\proof Follows from Lemma \ref{lemm:weak} since $p=-1$ and $q=1/2$ are conjugate numbers in $\left(-\infty,1\right]$. Note that we have omitted an additive constant in the objective functions of $(\mathcal{P})$ and $(\mathcal{D})$ in this section unlike Section \ref{intro}.\qed

We next show that having two feasible solutions $H$ and $u$ such that $f(H)=g(u)$ is not just sufficient but also necessary for optimality.
\BT \label{theo:opt_condTR} {\rm [Strong Duality]} There exist optimal solutions $H^*$ and $u^*$ for problems $(\P)$ and $(\D)$, respectively. Furthermore, the following conditions, together with primal and dual feasibility, are necessary and sufficient for optimality in both
$(\mathcal{P})$ and $(\mathcal{D})$: \begin{itemize}
\item[a.] $H^*=\frac{(M(u^*))^{-2}}{\trace{(M(u^*))^{-1}}}$,
\item[b.] $x_i^TH^*x_i=1$ if $u_i^*>0$.
\end{itemize} \ET
\proof As in the previous lemma, the proof follows from Theorem \ref{theo:opt_cond} for $p=-1$ and $q=1/2$.\qed

After some simplification, the necessary and sufficient conditions for $u^*$ to be optimal in $(\D)$ can be written as
\begin{itemize}
\item[(i)] $\alpha_i(u^*)\leq {u^*}^T\alpha(u^*)$ for all $i$, and 
\item[(ii)] $\alpha_i(u^*)={u^*}^T\alpha(u^*)$ if $u^*_i>0$,
\end{itemize} where $
  \alpha(u) := \nabla \hat{g}(u) = (x_i^T (M(u))^{-2} x_i)_{i=1}^m$. We say that a feasible solution $u$ for $(\mathcal{D})$ is \emph{$\epsilon$-primal feasible} if 
$\alpha_i(u)\leq u^T\alpha(u)(1+\epsilon)$ for all $i$, and say that it satisfies the \emph{$\epsilon$-approximate optimality conditions} or it is an \emph{$\epsilon$-approximate optimal solution} if moreover $\alpha_i(u)\geq u^T\alpha(u)(1-\epsilon)$ for all $i$ such that $u_i>0$. (Note that these definitions can be deduced from those in Section \ref{intro} for $p=-1$ and $q=1/2$.)

\BL \label{lem:dual_gapTR} Let $u$ be an $\epsilon$-primal feasible solution. Then we have \begin{itemize}\item[i.]  $0\leq g^*-g(u)\leq \ln
 (1+\epsilon)$
\item[ii.]  $1\leq \frac{\hat{g}(u)}{\hat{g}^*}\leq 1 +\epsilon,$
 \end{itemize} where $g^*$ and $\hat{g}^*$ are the optimal objective function values
 of $(\mathcal{D})$ and $(\hat{\mathcal{D}})$, respectively.\EL
\proof Since $u$ is an $\epsilon$-primal feasible solution, $\frac{(M(u))^{-2}}{(1+\epsilon)\trace{(M(u))^{-1}}}$ is feasible with respect to $(\mathcal{P})$. Let $H^*$ and $u^*$ be optimal solutions of $(\P)$
and $(\D)$, respectively. Then we have \bea -2\ln\trace\left(
\frac{(M(u))^{-2}}{(1+\epsilon)\trace{(M(u))^{-1}}}\right)^{1/2}+2\ln\trace H^{*1/2}&\geq&0, {\rm or}\nonumber\\
\ln(1+\epsilon)-\ln\trace (M(u))^{-1}-g(u^*)&\geq&0, {\rm from \ which} \nonumber\\
 0\leq g^*-g(u)\leq \ln(1+\epsilon),&& \eea which proves (i). Property (ii) follows from $g=-\ln(-\hat{g})$.\qed

%%%%%%%%%%%%%
\section{Algorithms and Analysis} \label{sec:TR}
%%%%%%%%%%%%%

In the rest of this paper, we will develop various iterative (Frank-Wolfe type) algorithms for solving $(\D)$ and $(\hat{\D})$. We will assume that the following assumption holds, for every feasible solution $u$ produced by these algorithms.

\BA\label{ass:upb} The dual feasible variable $u$ satisfies 
$\omega_j(u):=x_j^T(M(u))^{-1}x_j\leq \omega$ for all $j \in \{1,\dots,m\}$ and for some $\omega>1$.\EA

The objective function $\hat{g}$ of $(\hat{\mathcal{D}})$ is a concave function with gradient $\alpha(u)$
and that, with
\beq\label{eq:updateuTR}
  u_+ := (1-\tau) u + \tau e_j,
\eeq
rank-one update formulae give \bea \label{eq:imping}\hat{g}(u_+)&=& -\trace(M(u_+))^{-1}\nonumber\\
&=&-\trace\left((1+\lambda)\left((M(u))^{-1}-\frac{\lambda (M(u))^{-1}x_jx_j^T(M(u))^{-1}}{1+\lambda \omega_j(u)}\right)\right)\nonumber\\
&=&-(1+\lambda)\left(\trace(M(u))^{-1}-\frac{\lambda\trace\left((M(u))^{-1}x_jx_j^T(M(u))^{-1}\right)}{1+\lambda \omega_j(u)}\right)\nonumber\\
&=&(1+\lambda)\hat{g}(u)+\frac{\lambda(1+\lambda)}{1+\lambda \omega_j(u)}\alpha_j(u),\eea where $\lambda=\frac{\tau}{1-\tau}$. The partial derivative of the objective function is equal to \bea\frac{\partial \hat{g}(u_+)}{\partial \lambda}&=&\hat{g}(u)+\frac{\lambda^2\omega_j(u)+2\lambda+1}{(1+\lambda \omega_j(u))^2}\alpha_j(u).\eea Let $\hat{g}$, $\omega_j$, and $\alpha_j$ be shorthand for $\hat{g}(u)$, $\omega_j(u)$, and $\alpha_j(u)$, respectively. The numerator of the partial derivative is equal to the left-hand side of the following equation (the denominator is positive):
\bea \label{eq:TRroot}(\omega_j^2\hat{g}+\omega_j \alpha_j)\lambda^2+\lambda (2\omega_j \hat{g}+2\alpha_j)+\hat{g}+\alpha_j=0.\eea We can find the best step size $\tau^*$ (or $\lambda^*$) by investigating the roots of the quadratic equation (\ref{eq:TRroot}) and the boundary condition ($\lambda^*\geq-u_j$) arising from the nonnegativity of the dual feasible solutions as follows:
\begin{itemize}
\item if we have $\omega_j \hat{g}+\alpha_j = 0$, then the partial derivative is negative for all values and hence $\lambda^*=-u_j$;
\item if $(1 - \omega_j)(\alpha_j + \omega_j \hat g) < 0$  (which is equivalent to $\omega_j < 1$ since $\alpha_j + \omega_j \hat g\leq0$ for any feasible solution), the discriminant of the quadratic (\ref{eq:TRroot}) is negative. Furthermore,  $\hat{g} + \alpha_j < 0$ (since $0 \leq \omega_j < 1$, $\alpha_j + \omega_j \hat g < 0$, and $\hat{g} < 0$), and hence the quadratic (\ref{eq:TRroot}) has no real roots and everywhere negative. Therefore,  $\lambda^*=-u_j$;
\item otherwise $\lambda^*$ is equal to one of the roots of the quadratic (\ref{eq:TRroot}), which are \bea\lambda^*_{1,2}&=&\frac{-\omega_j \hat{g}-\alpha_j\pm\sqrt{(\omega_j \hat{g}+\alpha_j)^2-(\omega_j^2\hat{g}+\omega_j \alpha_j)(\hat{g}+\alpha_j)}}{(\omega_j^2\hat{g}+\omega_j \alpha_j)}\nonumber\\
&=& \frac{-\omega_j \hat{g}-\alpha_j\pm\sqrt{\alpha_j(1-\omega_j)(\alpha_j+\omega_j \hat{g})}}{(\omega_j^2\hat{g}+\omega_j \alpha_j)}\nonumber,\eea or $-u_j$ whichever is feasible and gives the greatest improvement in the objective function.
\end{itemize} 
Once we find the step size, we can calculate $\omega(u_+)$ and $\alpha(u_+)$ from \bea \label{update_xi}\omega_i(u_+)&=&x_i^T(M(u_+))^{-1}x_i\nonumber\\
&=&x_i^T\left((1+\lambda)\left((M(u))^{-1}-\frac{\lambda (M(u))^{-1}x_jx_j^T(M(u))^{-1}}{1+\lambda \omega_j(u)}\right)\right)x_i\nonumber\\
&=&(1+\lambda)\omega_i(u)-\frac{(1+\lambda)\lambda}{1+\lambda \omega_j(u)}\omega_{ij}(u)^2\nonumber\\
&=& (1+\lambda)(\omega_i(u)-\eta\omega_{ij}(u)^2),
\eea 
and 
\bea
\alpha_i(u_+)&=&x_i^T(M(u_+))^{-2}x_i\nonumber\\
&=& x_i^T((1+\lambda)\left((M(u))^{-1}-\frac{\lambda (M(u))^{-1}x_jx_j^T(M(u))^{-1}}{1+\lambda \omega_j(u)}\right)...\nonumber\\
&& (1+\lambda)\left((M(u))^{-1}-\frac{\lambda (M(u))^{-1}x_jx_j^T(M(u))^{-1}}{1+\lambda \omega_j(u)}\right))x_i\nonumber\\
&=& (1+\lambda)^2 x_i^T((M(u))^{-2}-\frac{2\lambda}{1+\lambda \omega_j(u)}(M(u))^{-2}x_jx_j^T(M(u))^{-1}...\nonumber\\
&&  +\frac{\lambda^2}{(1+\lambda \omega_j(u))^2}(M(u))^{-1}x_jx_j^T(M(u))^{-2}x_jx_j^T(M(u))^{-1})x_i\nonumber\\
&=& (1+\lambda)^2\alpha_i(u)-2\frac{(1+\lambda)^2\lambda}{1+\lambda \omega_j(u)}\omega_{ij}(u)\alpha_{ij}(u)+\frac{(1+\lambda)^2\lambda^2}{(1+\lambda \omega_j(u))^2}\omega_{ij}(u)^2\alpha_{j}(u)\nonumber\\
&=& (1+\lambda^2)(\alpha_i(u)-2\eta\omega_{ij}(u)\alpha_{ij}(u)+\eta^2\omega_{ij}(u)^2\alpha_j(u)), 
\eea 
where $\eta:=\frac{\lambda}{1+\lambda \omega_j(u)}$, 
$\omega_{ij}(u):=x_i^T(M(u))^{-1}x_j$, and 
$\alpha_{ij}(u):=x_i^T(M(u))^{-2}x_j$. Note that all updates can be performed cheaply 
(in $\mathcal{O}(nm)$ operations). 

Now we describe two Frank-Wolfe type algorithms. The first algorithm (Algorithm \ref{TR-KH}) uses positive step sizes and seeks an $\epsilon$-primal feasible solution; whereas the second one (Algorithm \ref{TR-TY}) may also have negative step sizes and stops when an $\epsilon$-approximate optimal solution is found. This algorithm is an extension of the first one with Wolfe's away steps. We will show that although these algorithms have similar global complexity results, away steps are necessary in order to achieve high accuracy, a phenomenon that is also observed for the D-Optimal Experimental Design Problem in \cite{AST08}.  

\begin{center}
\framebox{ \vbox{\hsize=1in
\BAL\label{TR-KH}\begin{tabbing}
\= \hspace*{.25in} \= \hspace*{.25in} \= \hspace*{.25in} \=
\hspace*{.25in} \=\kill 
\\\>\textbf{Input:} $X \in \R^{n \times m}$,
$\epsilon > 0$.
\\\> \textbf{Step 0.} Let $u=(1/m)e$. Compute $\omega(u)$ and $\alpha(u)$.\\
\>\textbf{Step 1.} Find $j := \arg\max_t \{\alpha_t(u) - u^T\alpha(u)\}$. \\
\>\>\textbf{If} $\frac{\alpha_j(u)}{u^T\alpha(u)}-1 \leq \epsilon$, \\
\>\>\>\textbf{STOP:} $u$ is an $\epsilon$-primal feasible solution.\\
\>\textbf{Step 2.} Replace $u$ as in (\ref{eq:updateuTR}), where
$\tau
> 0$
is chosen to maximize $\hat{g}$.\\
\>\textbf{Step 3.} Update $\omega(u)$ and $\alpha(u)$. \textbf{Go to} Step 1.
\end{tabbing}\EAL}
}
\end{center}

\begin{center}
\framebox{ \vbox{\hsize=1in
\BAL\label{TR-TY}\begin{tabbing}
\= \hspace*{.25in} \= \hspace*{.25in} \= \hspace*{.25in} \=
\hspace*{.25in} \=\kill 
\\\>\textbf{Input:} $X \in \R^{n \times m}$,
$\epsilon > 0$.
\\\> \textbf{Step 0.} Let $u=(1/m)e$. Compute $\omega(u)$ and $\alpha(u)$.\\
\>\textbf{Step 1.} Find $j := \arg\max_t \{\alpha_t(u) - u^T\alpha(u)\}$ and $i := \arg\min_t\{\alpha_t(u) - u^T\alpha(u):u_t>0\}$. \\
\>\>\textbf{If} $\frac{\alpha_j(u)}{u^T\alpha(u)}-1 \leq \epsilon$ and $1-\frac{\alpha_i(u)}{u^T\alpha(u)} \leq \epsilon$, \\
\>\>\>\textbf{STOP:} $u$ is an $\epsilon$-approximate optimal solution.\\
\>\>\textbf{Else,} \\
\>\>\>\textbf{if} $\alpha_j(u) - u^T\alpha(u) > u^T\alpha(u) - \alpha_i(u) $, \textbf{go to} Step 2;\\
\>\>\>\textbf{else,} \textbf{go to} Step 3. \\
\>\textbf{Step 2.} Replace $u$ as in (\ref{eq:updateuTR}), where
$\tau
> 0$
is chosen to maximize $g$. \textbf{Go to} Step 4.\\
\>\textbf{Step 3.} Replace $u$ by $u_+:= (1 - \tau)u + \tau e_i$,
where now $\tau$ is
chosen from \\ \>\>negative values to maximize $\hat{g}$ subject to $u_+$
remaining feasible. \\
\>\textbf{Step 4.} Update $\omega(u)$ and $\alpha(u)$. \textbf{Go to} Step 1.
\end{tabbing}\EAL}
}
\end{center}

If we look closely at these algorithms, we can identify three different types of iterations. Let $u^l$ be the dual feasible solution at hand at iteration number $l$, $e_{j_l}$ be the vertex that we use in our update and $\tau_l$ be the step size associated with this update. We refer to iteration $l$ as 
\begin{itemize}\item[-]an \emph{add/increase step} if $\tau_l>0$,
\item[-]a \emph{decrease step} if $u^{l}_{j_l}>0$ and $\frac{-u^{l}_{j_l}}{1-u^{l}_{j_l}}<\tau_l<0$, and
\item[-]a \emph{drop step} if $u^{l}_{j_l}>0$ and $\tau_l=\frac{-u^{l}_{j_l}}{1-u^{l}_{j_l}}$.\end{itemize}
We only have add/increase steps in Algorithm \ref{TR-KH}, whereas all types of steps can be performed in Algorithm \ref{TR-TY}. Note that after a drop step we have $u^{l+1}_{j_l}=0$. In such a step, we may not be able to improve the objective function as much as we desire. Fortunately, the number of drop steps is bounded above by the number of add steps plus a constant (the 
number of positive components of the initial solution), and hence studying only the
first two types of steps will be enough to obtain convergence results. 

\BL \label{lem:init_KHA_TR} $u^0=(1/m)e=\frac{1}{m}(1,1,\dots,1)$ is an $(m-1)$-primal feasible solution.\EL
\proof Follows from Lemma \ref{lem:dual_gap} in Section \ref{intro}. \qed

We now analyze the first algorithm closely:

\BL\label{lemm:CAlg1} As long as $u^l$ satisfy Assumption \ref{ass:upb} for all $l=1,2,\dots$, Algorithm \ref{TR-KH} finds an $\epsilon$-primal feasible solution in at most \bea
\label{{Alg1iter}}\mathcal{L}(\epsilon)=\mathcal{O}(\ln m+\epsilon^{-1})\eea steps. The constants hidden in the `big oh' are linearly dependent on the constant $\omega$ in Assumption \ref{ass:upb}.\EL
\proof Given a dual solution $u^l$ (the iterate at iteration $l$), we define $ \epsilon_l = \max \{\frac{\alpha_j(u^l)}{(u^l)^T\alpha(u^l)}-1 ,1-\frac{\alpha_i(u^l)}{(u^l)^T\alpha(u^l)} \}$, where $j := \arg\max_t \{\alpha_t(u^l) - (u^l)^T\alpha(u^l)\}$ and $i := \arg\min_t\{\alpha_t(u^l) - (u^l)^T\alpha(u^l):u^l_t>0\}$. (Note that the algorithm stops at iteration $k$ if $\epsilon_k \leq \epsilon$.)

We will first prove that \bea \label{L1}\mathcal{L}(1)=\min\{l|\epsilon_l\leq1\}=\mathcal{O}(\ln m).\eea Let $j_l$ be the index of the pivot point at iteration $l$, $\tau_l$ be the step size, and $\lambda_l=\frac{\tau_l}{1-\tau_l}$. (Remember that all values of $\hat{g}$ are negative by definition.) At each iteration $l$ with $\epsilon_l\geq1$, from (\ref{eq:imping}), we have 
\bea \hat{g}(u^{l+1})-\hat{g}(u^l)& = & \lambda_l\hat{g}(u^l)+\frac{\lambda_l(1+\lambda_l)}{1+\lambda_l\omega_{j_l}(u^l)}\alpha_{s_l}\nonumber\\
& \geq & \frac{1}{2\omega_{j_l}(u^l)}\hat{g}(u^l)-\frac{\frac{1}{2\omega_{j_l}(u^l)}}{1+\frac{1}{2\omega_{j_l}(u^l)}\omega_{j_l}(u^l)}2\hat{g}(u^l)\nonumber\\
& \geq & \frac{1}{2\omega_{j_l}(u^l)}\hat{g}(u^l)\left(1-\frac{2}{1+\frac{1}{2}}\right)\nonumber\\
& \geq & -\frac{\hat{g}(u^l)}{6\omega}.
\eea The first inequality follows since the improvement obtained from choosing the best step length is at least as good as the improvement obtained by using any step length; in particular, it can be bounded by plugging in $\lambda_l=\frac{1}{2\omega_{j_l}(u^l)}$. 

Hence we have \bea\label{ggu} \hat{g}(u^{l+1})\geq (1-\frac{1}{6\omega})\hat{g}(u^l). \eea Using Lemmas \ref{lem:dual_gapTR} and \ref{lem:init_KHA_TR}, \bea \label{gg0}\hat{g}(u^0)\geq m\hat{g}^*.\eea Combining inequalities (\ref{ggu}) and (\ref{gg0}), we obtain \bea \hat{g}^*\geq \hat{g}(u^l)\geq & (1-\frac{1}{6\omega})^l\hat{g}(u^0)\geq (1-\frac{1}{6\omega})^lm\hat{g}^*\geq e^{-\frac{l}{6\omega}}m\hat{g}^*.\eea Hence we must have $\mathcal{L}(1)\leq 6\omega \ln(m)=\mathcal{O}(\ln m).$

Now assume that $\epsilon_l\leq 1$ and define $h(\epsilon_l):=\min\{h|\epsilon_{l+h}\leq\epsilon_l/2\}$. As long as $\epsilon_{l+h}\geq\epsilon/2$, from (\ref{eq:imping}) we also have 
\bea \label{impsmall}\hat{g}(u^{l+h+1})-\hat{g}(u^{l+h}) &\geq &\hat{g}(u^{l+h})\frac{\epsilon_l}{4\omega_{l+h}(u^l)}\left(1-\frac{1+\epsilon_l/2}{1+\frac{\epsilon_l}{4\omega_{l+h}(u^l)}\omega_{l+h}(u^l)}\right)\nonumber\\ &\geq & -\frac{\epsilon_l^2}{32\omega}\hat{g}^*.\eea Again, the first inequality is obtained by setting $\lambda_l=\frac{\epsilon_l}{4\omega_{l+h}(u^l)}$. On the other hand, Lemma \ref{lem:dual_gapTR} gives\bea \label{gapsmall}\frac{\hat{g}(u^l)}{\hat{g}^*}\leq {1+\epsilon_l}.\eea Combining equations (\ref{impsmall}) and (\ref{gapsmall}), we get $h(\epsilon_l)\leq \frac{32\omega}{\epsilon_l}$. Therefore \bea \label{Leps} \mathcal{H}(\epsilon)&=&h(\epsilon_l)+h(\epsilon_l/2)+h(\epsilon_l/4)+\dots+h(\epsilon_l/2^{\left\lceil \ln{\epsilon_l}/\epsilon\right\rceil-1})\nonumber\\
&\leq & 32\omega\left(\frac{1}{\epsilon_l}+\frac{2}{\epsilon_l}+\frac{4}{\epsilon_l}+\dots+\frac{2^{\left\lceil \ln{\epsilon_l}/\epsilon\right\rceil-1}}{\epsilon_l}\right)\leq\frac{64\omega}{\epsilon}=\mathcal{O}(\epsilon^{-1}),\eea iterations are required to obtain an $\epsilon$-primal feasible solution starting with a solution $\epsilon_l\leq 1$. Combining (\ref{Leps}) and (\ref{L1}) completes the proof. \qed

Once we take care of the drop steps, the analysis of the algorithm with away steps is no more complicated.

\BL\label{lemm:CAlg2} As long as $u^l$ satisfy Assumption \ref{ass:upb} for all $l=1,2,\dots$, Algorithm \ref{TR-TY} finds an $\epsilon$-approximate optimal solution in at most \bea\label{Alg2iter}\mathcal{L}(\epsilon)=\mathcal{O}(m+\epsilon^{-1})\eea steps. The constants hidden in the `big oh' are linearly dependent on the constant $\omega$ in Assumption \ref{ass:upb}.\EL
\proof We can only have add/increase steps when $\epsilon_l\geq 1$; hence Algorithms \ref{TR-KH} and \ref{TR-TY} take the same steps until the first solution $u_{\hat{l}}$ with $\epsilon_{\hat{l}}\leq 1$ is encountered. So that \bea \label{L2}\mathcal{L}(1)=\min\{l|\epsilon_l\leq1\}=\mathcal{O}(\ln m) \eea holds for Algorithm \ref{TR-TY} as well.  

Now assume that $\epsilon_l\leq 1$ and define $h(\epsilon_l):=\min\{h|\epsilon_{l+h}\leq\epsilon_l/2\}$ as before. Let us look at the improvement in the objective function at the $(l+h)^{th}$ iteration. There are three cases:
\begin{enumerate}
\item If this is an add/increase step, then 
\bea 
\hat{g}(u^{l+h+1})-\hat{g}(u^{l+h})&\geq & -\frac{\epsilon_l^2}{32\omega}\hat{g}^*
\eea 
from (\ref{impsmall});
\item if it is a decrease step, we have 
\bea 
\hat{g}(u^{l+h+1})-\hat{g}(u^{l+h})&\geq &\hat{g}(u^{l+h})\frac{-\epsilon_l}{4\omega_{l+h}}\left(1-\frac{1-\epsilon_l/2}{1-\frac{\epsilon_l}{4\omega_{l+h}}\omega_{l+h}}\right)\nonumber\\ 
&\geq & -\frac{\epsilon_l^2}{16\omega}\hat{g}^*;
\eea
\item otherwise (it is a drop step), we can only conclude that 
\bea 
\hat{g}(u^{l+h+1})-\hat{g}(u^{l+h})&\geq 0.
\eea 
\end{enumerate}
Hence we have \bea \label{impsmall2} \hat{g}(u^{l+h+1})-\hat{g}(u^{l+h})&\geq & -\frac{\epsilon_l^2}{32\omega}g^*,\eea whenever we have an add/increase or decrease step. 

On the other hand, using Lemma \ref{lem:dual_gapTR} we have\bea \label{gapsmall2}\frac{\hat{g}(u^l)}{\hat{g}^*}\leq {1+\epsilon_l}.\eea 
Combining equations (\ref{impsmall2}) and (\ref{gapsmall2}), we need to perform at most
$$h(\epsilon_l)\leq \frac{32\omega}{\epsilon_l}$$ add/increase and decrease steps to obtain an $\epsilon_l/2$-approximate optimal solution starting with an $\epsilon_l$-approximate optimal solution. Applying this argument repeatedly, we conclude that we need at most \bea \mathcal{H}(\epsilon)&=&h(\epsilon_l)+h(\epsilon_l/2)+h(\epsilon_l/4)+\dots+h(\epsilon_l/2^{\left\lceil \ln{\epsilon_l}/\epsilon\right\rceil-1})\nonumber\\
&\leq & 32\omega\left(\frac{1}{\epsilon_l}+\frac{2}{\epsilon_l}+\frac{4}{\epsilon_l}+\dots+\frac{2^{\left\lceil \ln{\epsilon_l}/\epsilon\right\rceil-1}}{\epsilon_l}\right)\nonumber\\&\leq&\frac{64\omega}{\epsilon}=\mathcal{O}(\epsilon^{-1} ),\eea add/increase and decrease iterations to obtain an $\epsilon$-approximate optimal solution starting with an $\epsilon_l$-approximate optimal solution where $\epsilon_l\in(0,1]$. Since the number of drop steps is bounded above by the number of add steps plus $m$ (the number of positive components of the initial solution $u^0$), (\ref{Alg2iter}) is immediate.\qed

The following lemma shows that (for the same set of data points) an approximate solution to the D-optimal design problem is also close to the optimal solution of the A-optimal design problem in some sense.

\BL \label{lem:init_gapKY} Let $u^D$ be a $\delta$-primal feasible solution for the D-optimal design (as defined as in Definition \ref{def:sol} with $p=q=0$), then $u^D$ is an  $(n+n\delta-1)$-primal feasible solution for $(\mathcal{D})$.\EL
\proof For all $1\leq j \leq m$, we have \beann 
x_j^T(M(u^D)X^T)^{-2}x_j&=&\trace((M(u^D)X^T)^{-1}(M(u^D)X^T)^{-1/2}x_jx_j^T(M(u^D)X^T)^{-1/2})\\ 
&\leq&\trace((M(u^D)X^T)^{-1})\trace((M(u^D)X^T)^{-1/2}x_jx_j^T(M(u^D)X^T)^{-1/2})\\ 
&\leq&\trace((M(u^D)X^T)^{-1})\trace(x_j^T(M(u^D)X^T)^{-1}x_j)\\
&\leq&\trace((M(u^D)X^T)^{-1})(n+n\delta),\eeann
where $U^D=\Diag(u^D)$. This proves that $u^D$ is an $(n+n\delta-1)$-primal feasible solution for $(\mathcal{D})$.\qed

Let us call the algorithm which finds a 1-approximate optimal solution for the D-optimal design problem using WA-TY method described in \cite{TYil07} and proceeds with Steps 1, 2, and 3 of Algorithm \ref{TR-KH} as Algorithm \ref{TR-KH}-MV; and that proceeds with Steps 1, 2, 3, and 4 of Algorithm \ref{TR-TY} as Algorithm \ref{TR-TY}-MV. When $m\gg n$, these algorithms perform significantly better than the original ones as the following lemma suggests. In addition, we are able to obtain core-set results for free.

\BL \label{lemm:CAlgD} As long as $u^l$ satisfy Assumption \ref{ass:upb} for all $l=1,2,\dots$, 
\begin{itemize}
\item[a.] Algorithm \ref{TR-KH}-MV finds an $\epsilon$-primal feasible solution in at most \bea\label{Alg1Diter}\mathcal{L}(\epsilon)=\mathcal{O}(n\ln n+\epsilon^{-1})\eea steps;
\item[b.] Algorithm \ref{TR-TY}-MV finds an $\epsilon$-approximate optimal solution in at most \bea\label{Alg2Diter}\mathcal{L}(\epsilon)=\mathcal{O}(\min\{m, n\ln n\}+\epsilon^{-1})\eea steps; 
\item[c.] furthermore, Algorithm \ref{TR-KH}-MV identifies a set $\mathcal{A}\subset\mathcal{X}$ such that 
$$|\mathcal{A}|\leq\mathcal{O}(n\ln n+\epsilon^{-1})$$
and an $\epsilon$-primal feasible solution $u$ for the A-optimal design problem defined over data set $\mathcal{A}$ is also an $\epsilon$-primal feasible solution for the A-optimal design problem defined over data set $\mathcal{X}$; and 
\item[d.] Algorithm \ref{TR-TY}-MV identifies a set $\mathcal{A}\subset\mathcal{X}$ such that 
$$|\mathcal{A}|\leq\mathcal{O}(n\ln n+\epsilon^{-1})$$
and an $\epsilon$-approximate optimal solution $u$ for the A-optimal design problem defined over data set $\mathcal{A}$ is also an $\epsilon$-approximate optimal solution for the A-optimal design problem defined over data set $\mathcal{X}$.
\end{itemize}\EL
\proof It is proved in \cite{TYil07} that a 1-approximate optimal solution for the D-optimal design problem can be obtained in $\mathcal{O}(n\ln n)$ iterations. Let $u^0$ be such a solution. Lemmas \ref{lem:dual_gapTR} and \ref{lem:init_gapKY} give \bea \label{ggTY}\hat{g}(u^0)\geq 2n\hat{g}^*.\eea Replacing (\ref{gg0}) with (\ref{ggTY}) in the proof of Lemma \ref{lemm:CAlg1}, gives $\mathcal{L}(1)=\mathcal{O}(\ln n)$ for Algorithm \ref{TR-KH}-MV. Since the rest of the proof is unchanged, Algorithm \ref{TR-KH}-MV finds an $\epsilon$-primal feasible solution in $\mathcal{L}(\epsilon)=\mathcal{O}(n\ln n+\ln n +\epsilon^{-1})=\mathcal{O}(n\ln n+\epsilon^{-1})$ iterations, which proves (a).

Similarly, (b) follows from Lemma \ref{lemm:CAlg2} with replacing $\mathcal{L}(1)=\mathcal{O}(\ln n)$ and noticing that the number of positive components in $u^0$ is bounded above by $\mathcal{O}(\min\{m, n\ln n\})$ as proved in \cite{KumYil05}. 

Let $\hat{u}$ be the output of Algorithm \ref{TR-KH}-MV. Letting $\mathcal{A}=\{x_i: \hat{u}_i>0\}$ proves (c) since the number of positive components of $\hat{u}$ is bounded above by the number of positive components in the initial solution (which is $2n$ as discussed in \cite{KumYil05}) plus the number of add steps (which is less than the total number of iterations proved in part (a)). Similar arguments can be used to prove part (d).\qed

\BR The complexity results we have presented in this section depend on the constant $\omega$ in Assumption \ref{ass:upb}. It is easy to see that $\omega$ is proportional to the inverse of the infimum of the set of eigenvalues of the matrices $M(u^l)$, $l = 1,2, \dots$ generated by the algorithm.  Alternatively, $\omega$ is the supremum of the ellipsoidal distances of the data points with respect to the ellipsoids centered at the origin that have shape matrices $M(u^l)$.  Therefore, $\omega$ depends on the geometry of the design points and the steps taken by the algorithm. When the design points are very thinly spread around a proper subspace of $\R^n$, the ellipsoids generated by the algorithm will have elongated axes in some directions and extremely short axes in others, potentially leading to large ellipsoidal distances for some data points. The sequence of positive definite matrices, $M(u^l)$, $l = 1,2, \dots$, generated by the algorithm  converge to a single limit point, say $M(u^*)$, which is positive definite and has smallest eigenvalue, say $\lambda^*$. Therefore, there exists an integer $N(\frac{\lambda^*}{2})$, such that the eigenvalues of $M(u^l)$, for $l\geq N\left(\frac{\lambda^*}{2}\right)$ are lower bounded by $\lambda^*/2$. That guarantees that $\omega$ is finite.
\ER

%%%%%%%%%%%%%%%

\section{Local Convergence Properties}\label{sec:TRloc}

\begin{figure}
		\begin{tabular}{ll}
			\centering
        \includegraphics[width=3in]{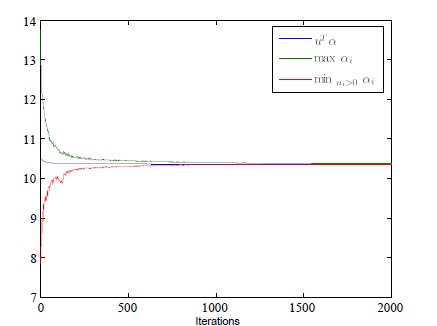}
    %\label{fig:fig2a}
&
%\caption{Average number of iterations to obtain a $10^{-4}$-approximate optimal solution using Algorithm 2 for randomly generated data sets with $(n,m) = (100,10000)$ and various values of $k$}
    \centering
        \includegraphics[width=3in]{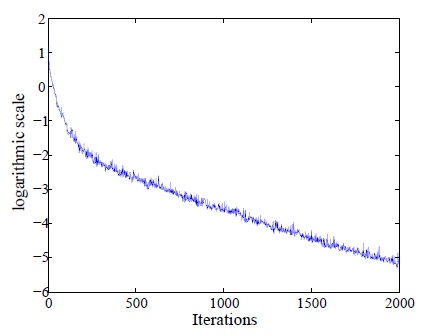}
		\end{tabular}
\caption{Behavior of Algorithm \ref{TR-TY} for $(m,n)=(10000,100)$.}
 \label{fig:figlinTR} \end{figure}

%%%%%%%%%%%%%%%
In this section, we will show that Algorithms \ref{TR-TY} and \ref{TR-TY}-MV are locally linearly convergent, i.e., the number of iterations grows with $\mathcal{O}(\ln\epsilon^{-1})$ not $\mathcal{O}(\epsilon^{-1})$ asymptotically under certain assumptions. The typical behavior of the algorithms is demonstrated in Figure \ref{fig:figlinTR}. Unfortunately, this bound depends on the data of the problem as well as the dimensions and the constant $\omega$ defined as in Lemma \ref{lemm:CAlg2}, and so does not provide global complexity bounds better than those above. 
Let us look at the following perturbation of the primal problem ($\mathcal{P}$):
$$\ba{rrrcl}
  & \min & f(H) &:=& - 2\ln \trace H^{1/2} \\
  (\mathcal{P}(\kappa))   &           & x_i^T H x_i & \leq & 1+\kappa_i, \, i =
  1,\dots,m.
\ea $$ Given $u$ satisfying the $\epsilon$-approximate optimality conditions, let $H(u):=\frac{(M(u))^{-2}}{\trace(M(u))^{-1}}$ and define $\kappa:=\kappa(u,\epsilon)$ as 

\[
  \kappa_i(u,\epsilon) := \left\{ \ba{ll} \epsilon & \mbox{if } u_i = 0, \\
                         x_i^T H(u) x_i - 1 & \mbox{else.} \ea \right.
\] Note that, each component of perturbation vector $\kappa$ is absolutely bounded by $\epsilon$ and $u^T\kappa=\frac{\sum_{j:u_j>0}u_jx_j^T(M(u))^{-2}x_j}{\trace(M(u))^{-1}}-1=1-1=0$. $H(u)$ is optimal w.r.t. $\mathcal{P}(\kappa(u,\epsilon))$, since it is feasible and $u$ provides the corresponding Lagrangian multipliers. Let $\phi(\kappa)$ be the value function, the optimal value of ($\mathcal{P}(\kappa)$). If $u^*$ is a vector of multipliers corresponding to the optimal solution of ($\mathcal{P}$), then $u^*$ is a subgradient of $\phi$ at 0. For any $\epsilon$-approximate optimal solution $u$ and $\kappa:=\kappa(u,\epsilon)$, we have \bea g(u)=f(H(u))&=&\phi(\kappa)\geq\phi(0)+u^{*^T}\kappa\nonumber\\
&=&g^*+(u^*-u)^T\kappa\geq g^*-\|u-u^*\|\|\kappa\|.\eea
Since $f(H)$ is strongly convex near any $H\succ0$ and the constraints are linear, Robinson's second order condition holds at $(H,\hat{u})$ for any $\mathcal{P}(\kappa)$, where $H$ is the optimal solution and $\hat{u}$ is any Lagrangian multiplier. Moreover, the linear constraints are regular at any feasible point and they are polyhedral, therefore Robinson's Corollary 4.3 (\cite{Rob82}) applies, which shows that 
$$\|u-u^*\|\leq L\|\kappa\|\leq L\sqrt{m}\epsilon,$$ where $L$ is a data-dependent constant and
whenever $\|\kappa\|$ is sufficiently small. Hence we conclude \bea\label{gap}g^*-g(u)\leq M\epsilon^2\eea
for some $M$ depending on the data of the problem ($\mathcal{P}$). Using inequality (\ref{gap}), we can find a constant $\hat{c}$ such that \bea \label{gap2} \frac{\hat{g}(u^l)}{\hat{g}^*}\leq e^{M\epsilon_l^2}\leq 1+\hat{c}\epsilon_l^2,\eea for any $\epsilon_l$-approximate solution $u^l$, as long as $\epsilon_l$ is small enough. Using (\ref{gap2}) instead of (\ref{gapsmall2}) in the last part of the proof of Lemma \ref{lemm:CAlg2} we obtain the following lemma: 

\BL \label{local22} Under the assumption of Lemma \ref{lemm:CAlg2}, there exists a data-dependent constant $Q$ such that Algorithms \ref{TR-TY} and \ref{TR-TY}-MV discussed above converges to an $\epsilon$-approximate optimal solution in $\O(Q+\ln(1/\epsilon))$ steps. \EL

%%%%%%%%%%%%%
\section{Computational Study} \label{sec:TRcomp}
%%%%%%%%%%%%%

In this section we present some computational results for Algorithms \ref{TR-KH} and \ref{TR-TY}, using different initialization strategies: the Khachiyan initialization (KH) strategy, where the initial feasible solution $u$ is the center of the unit simplex, i.e., $u_i=1/m$ for all $i=1,\dots,m$; the Kumar-Y{\i}ld{\i}r{\i}m initialization (KY) strategy introduced in \cite{KumYil05}; and a new strategy (MV) where the initial solution is set to be a 1-approximate optimal solution obtained by the WA-TY method of \cite{TYil07}. All experiments were carried out on a 3.40 GHz Pentium
IV processor with 1.0 GB RAM using MATLAB version R2006b. We assume a general linear model in this section and next. Note that we do not generate our regression points as fixed grid of support points on a compact interval as many other papers do. Instead we generate a large set of random regression points following \cite{SunFre02}. According to our past experience from \cite{AST08}, instances generated by this method are quite challenging. For all algorithms we study below, we report the total computational time inclusive of the time spent on the initialization schemes.

In Table \ref{comp_1TR}, we compare the computation time of the algorithms described above with three initializations on small- to medium-sized data sets. The data sets are generated as in \cite{SunFre02}. The results presented are the geometric means of the solution times for 10 random problems to obtain an $\eps$-primal feasible (for Algorithm \ref{TR-KH}) or an $\eps$-approximate optimal solution (for Algorithm \ref{TR-TY}) where $\eps=10^{-3}$. It is clear from the results that Algorithm \ref{TR-TY} preforms significantly better than Algorithm \ref{TR-KH} showing that away steps are necessary for developing efficient algorithms. For these instances, it is hard to make conclusions on the performances of the initialization strategies. 

\begin{table}[ht]
\begin{center}
\caption{Geometric mean of solution times of Algorithms \ref{TR-KH} and \ref{TR-TY} for small-medium sized problems with different initializations}\label{comp_1TR} \vspace{.1in}
\begin{tabular}{|c|r||r|r|r||r|r|r|}\hline
 & & \multicolumn{6}{|c|}{Geometric Mean of Time (Seconds)} \\\cline{3-8} 
& & \multicolumn{3}{|c||}{Algorithm \ref{TR-KH}}&\multicolumn{3}{|c|}{Algorithm \ref{TR-TY}} \\
 \cline{3-8} n   &   m  &   Kha &KY  & MV & Kha & KY &MV   \\ \hline
10	&	50		& 9.1		& 8.5		& 8.5		& 1.6	& 0.7	& 0.8	\\
10	&	100		& 10.5	& 10.3	& 10.1	& 1.2	& 1.3	& 1.9	\\
10	& 200		& 10.8	& 9.9		& 10.6	& 0.6	& 1.4	& 1.1	\\
10	& 400		& 11.9	& 11.2	& 12.5	& 0.4	& 0.8	& 1.0	\\
10	& 600		& 13.3	& 13.0	& 12.7	& 0.6	& 1.1	& 0.8	\\
10	& 800		& 13.9	& 13.4	& 13.4	& 1.0	& 1.5	& 1.2	\\
20	& 200		& 37.9	& 36.4	& 35.3	& 1.2	& 0.8	& 0.6	\\
20	& 300		& 39.6	& 40.0	& 39.2	& 1.4	& 1.1	& 1.0	\\
20	& 400		& 38.3	& 38.5	& 39.7	& 0.7	& 1.7	& 1.6	\\
20	& 600		& 49.2	& 49.2	& 45.7 	& 0.9	& 2.0	& 2.9	\\
20	& 800		& 52.6	& 54.5	& 52.3	& 1.2	& 2.5	& 3.4	\\
20	& 1000	& 57.1	& 54.4	& 53.1	& 1.7	& 3.4	& 3.4	\\
20	& 1200	& 58.7	& 56.4	& 56.6	& 1.8	& 5.3	& 5.0	\\
30	& 450		& 108.6	& 100.1	& 93.9	& 2.0	& 2.9	& 2.8	\\
30	& 900		& 130.0	& 119.6	& 127.5	& 1.5	& 4.7	& 4.5	\\
30	& 1350	& 142.3	& 121.3	& 120.9	& 2.3	& 6.5	& 5.8	\\
30	& 1800	& 154.2	& 131.3	& 128.9	& 3.5	& 7.6	& 7.7	\\
  \hline
\end{tabular}\end{center}
\end{table}

\begin{table}[ht]
\begin{center}
\caption{Geometric mean of solution times of Algorithms \ref{TR-KH} and \ref{TR-TY} for large problems with different initializations}\label{comp_2TR} \vspace{.1in}
\begin{tabular}{|c|r||r|r|r||r|r|r|}\hline
 & & \multicolumn{6}{|c|}{Geometric Mean of Time (Seconds)} \\\cline{3-8} 
& & \multicolumn{3}{|c||}{Algorithm \ref{TR-KH}}&\multicolumn{3}{|c|}{Algorithm \ref{TR-TY}} \\
 \cline{3-8} n   &   m  &    Kha &KY  & MV & Kha & KY &MV   \\ \hline

5		&	10000	&	17.267	&	12.208	&	11.641	&	35.236	&	3.5327	&	3.5428	\\
5		&	20000	&	26.57		&	20.417	&	20.905	&	55.491	&	7.8292	&	7.4747	\\
5		&	30000	&	35.941	&	29.808	&	30.374	&	43.136	&	7.9607	&	9.8677	\\
5		&	50000	&	58.433	&	54.698	&	52.828	&	98.456	&	28.159	&	28.715	\\
10	&	10000	&	43.677	&	32.431	&	32.173	&	38.017	&	5.7187	&	5.5486	\\
10	&	20000	&	76.886	&	67.377	&	66.554	&	138.93	&	10.604	&	10.154	\\
10	&	30000	&	103.56	&	87.166	&	90.091	&	126.69	&	17.158	&	15.499	\\
20	&	10000	&	141.76	&	113.23	&	117.45	&	48.849	&	18.482	&	19.234	\\
20	&	20000	&	211.44	&	186.48	&	183.35	&	196.31	&	40.659	&	39.256	\\	
20	&	30000	&	287.15	&	253.81	&	252.65	&	385.37	&	53.223	&	45.749	\\
20	&	50000	&	426.9		&	395.6		&	402.68	&	543.22	&	99.232	&	91.305	\\
30	&	10000	&	295.09	&	247.77	&	243.47	&	59.061	&	27.439	&	31.508	\\
30	&	20000	&	451.68	&	395.66	&	402.26	&	220.01	&	74.113	&	61.231	\\
30	&	30000	&	606.04	&	536.8		&	528.98	&	500.77	&	89.2		&	96.194	\\
50	&	50000	&	2308.2	&	2154.5	&	2142.8	&	1992.3	&	370.77	&	327.79	\\
 \hline
\end{tabular}\end{center}
\end{table}

\begin{table}[ht]
\begin{center}
\caption{Geometric mean of solution times of Algorithm \ref{TR-TY}-MV with different (update) selection strategies for small instances}\label{comp_3TR} \vspace{.1in}
\begin{tabular}{|c|r||r|r||r|r|}\hline
& & \multicolumn{2}{|c||}{Time (Seconds)}&\multicolumn{2}{|c|}{Iterations} \\
 \cline{3-6} n   &   m  & ALL &    Orig. & ALL &    Orig.  \\ \hline
20	&	200	&	0.54	&	0.85	&	510.7	&	1697.9\\
20	&	300	&	0.67	&	1.16		&	638.5	&	2252\\
20	&	400	&	0.91	&	1.72	&	772.08	&	3122\\
20	&	600	&	1.45		&	2.02	&	904.9	&	3254\\
20	&	800	&	2.01	&	2.57	&	1028.9	&	3918.6\\
20	&	1000	&	2.67	&	3.41	&	1189.9	&	4836.6\\
20	&	1200	&	3.00	&	5.35	&	1195.3	&	6397\\
30	&	450		&	1.26	&	2.90	&	963.3	&	4467.3\\
30	&	900		&	2.82	&	4.71	&	1314.6	&	5723.7\\
30	&	1350	&	4.68	&	6.59	&	1660.4	&	6976.3\\
30	&	1800	&	6.33	&	7.67	&	1782.9	&	7706.8\\
20	&	1000	&	2.54	&	3.49	&	1168.4	&	4694.9\\
 \hline
\end{tabular}\end{center}
\end{table}

\begin{table}[ht]
\begin{center}
\caption{Geometric mean of solution times of Algorithm \ref{TR-TY}-MV with different (update) selection strategies for large instances}\label{comp_4TR} \vspace{.1in}
\begin{tabular}{|c|r||r|r||r|r|}\hline
& & \multicolumn{2}{|c||}{Time (Seconds)}&\multicolumn{2}{|c|}{Iterations} \\
 \cline{3-6} n   &   m  & ALL &    Orig. & ALL &    Orig.  \\ \hline
10	&	10000	&	13.33		&	5.71	&	875.8	&	2656.5\\
20	&	10000	&	26.08	&	18.48	&	1634.5	&	6072.5\\
20	&	20000	&	59.32	&	40.61	&	1879.8	&	6852.7\\
20	&	30000	&	102.14	&	62.41	&	2220.3	&	7854.7\\
30	&	10000	&	42.95	&	27.43	&	2547.9	&	7100.8\\
30	&	20000	&	101.86	&	74.11	&	3085.6	&	10515\\
30	&	30000	&	140.42	&	89.2		&	2876.5	&	8899.2\\
50	&	50000	&	428.3		&	370.7	&	5106.4	&	15979\\
 \hline
\end{tabular}\end{center}
\end{table}

Table \ref{comp_2TR} presents the performance of the algorithms on larger data sets. Again, the results are the geometric means of the solution times of 10 random problems generated as in \cite{SunFre02} for each parameter set. The results indicate that for these instances where $m\gg n$, the MV initialization is outperforming the Khachiyan initialization as Lemma \ref{lemm:CAlgD} suggests. Since the KY initialization is somehow close to the MV initialization, its performance it similar to the MV initialization. One should not be surprised by the fact that Algorithm \ref{TR-TY} with the Khachiyan initialization is very slow on these instances, since the initial solution has many entries with positive weights and the algorithm needs to take many drop steps before converging to the optimal solution. Fortunately, other two initializations are able to find accurate solutions in short time. We have tried even larger data sets to explore the limits of the algorithms. We were able to find $10^{-4}$-approximate optimal solutions to instances where $n=500$ and $m=10000$ (generated as before) with Algorithm \ref{TR-TY} using KY initialization under 30 minutes. 

The number of iterations required can be significantly decreased if we make the best possible update (not just one of the two arguments used in Step 1) at each iteration. This can be done by calculating the improvement related to each index and choosing the best. We have coded a version of Algorithm \ref{TR-TY}-MV and experimented on some of the data sets above. The (mean) solution times and number of iterations are compared in Tables \ref{comp_3TR} and \ref{comp_4TR}. The unmodified version of the algorithm is represented in the columns labeled with `'Orig.`' while the version with optimal decisions is labeled with `'ALL'`. It is obvious that as the number of points in the data set increase calculating the possible improvement for each index becomes expensive; hence considering only two promising vertices is a wise choice. Obviously some hybrid versions, which choose the best of a small set of carefully selected indices, can perform better for certain instances; so can other versions with active set strategies.

%%%%%%%%%%%%%
\section{Semidefinite Programming Reformulation and Comparison} \label{sec:TRcompSDP}
%%%%%%%%%%%%%

Any reader with some familiarity with nonlinear optimization would know that semidefinite programming has gained significant attention in last two decades. As discussed in \cite{WSV00}, many interesting problems in science and engineering can be reformulated as SDPs and solved via one of the freely available SDP solvers such as SDPT3 or SEDUMI. The D-optimal and A-optimal design problems are no exception. Section 4 of \cite{VB99} provides the reformulations of both of these problems. Following their discussion, problem $(\mathcal{D})$ is equivalent to:

$$\ba{cccl}
   \min & \sum_{i=1}^n{t_i} \\
  (\mathcal{SDP})  &    \quad    \left(\ba{cc} M(u) & e_i \\ e_i^T & t_i   \ea \right) \succeq 0 , \, i =
  1,\dots,n,\\
 & e^Tu = 1,\\
& u \geq 0,
\ea$$ 
where $e_i$ is the $i^{th}$ unit vector in $\R^n$, and the variables are $u \in \R^m$ and $t \in \R^n$.

Although many problems can be cast as semidefinite programs, not many semidefinite formulations can be solved efficiently yet due to high memory requirements and slow convergence rate of the state-of-the-art methods. We compare one of our algorithms (Algorithm \ref{TR-TY} with KY initialization) versus the SDPT3 algorithm using the CVX platform on MATLAB, which is a classic platform to solve SDPs. The results presented in Table \ref{comp_SDP} are mean solution times for 5 random problems to obtain an $\eps$-approximate optimal solution with Algorithm \ref{TR-TY} in the third column and with the SDP solver on the forth. For fair comparison, we run both algorithms until a very accurate solution is obtained (i.e., $\eps=10^{-7}$), especially since being able  find accurate solutions is one of the strong points of the SDP approach. In this section, we test only 5 instances of each problem since the SDP solver takes very long amount of time and the conclusion is obvious even with small number of instances considered. The instances are generated as before following \cite{SunFre02}. The solutions obtained from the two methods were identical (to be precise: the norm of their distance was smaller than $10^{-7}$ as expected). It is clear that our first-order technique dominates the SDP method, sometimes it is more than 300 times faster. Furthermore, it is impossible to solve large instances of the SDP formulation due to memory restrictions and time limitations. For example, we can not solve problems with $n=30$ and $m=600$ with the SDP solver.

\begin{table}[ht]
\begin{center}
\caption{Mean solution times of SDPT3 and Algorithm 2-MV for small-medium sized problems}\label{comp_SDP} \vspace{.1in}
\begin{tabular}{|c|c|c|c||c|}\hline
 n   &   m  & Algorithm \ref{TR-TY}& SDP & speed-up   \\ \hline
10	&	50	&	0.33	&	0.58	&	3.20	\\
10	&	100	&	0.24	&	0.45	&	3.63	\\
10	&	200	&	0.70	&	1.51	&	1.90	\\
10	&	400	&	0.76	&	2.80	&	5.32	\\
10	&	600	&	1.56	&	7.83	&	3.01	\\
10	&	800	&	1.37	&	9.90	&	12.18	\\
10	&	1000	&	0.88	&	11.43	&	7.85	\\
20	&	50	&	0.08	&	2.66	&	42.40	\\
20	&	100	&	0.33	&	4.83	&	8.20	\\
20	&	200	&	0.44	&	11.40	&	19.64	\\
20	&	400	&	0.92	&	33.46	&	55.77	\\
20	&	600	&	1.15	&	66.73	&	54.49	\\
20	&	800	&	2.02	&	120.82	&	66.19	\\
20	&	1000	&	1.84	&	197.45	&	115.66	\\
30	&	50	&	0.05	&	17.07	&	334.33	\\
30	&	100	&	0.19	&	22.36	&	145.58	\\
30	&	200	&	0.89	&	58.38	&	38.88	\\
30	&	400	&	1.07	&	140.58	&	187.42	\\
30	&	600	&	2.13	&	337.95	&	139.66	\\

  \hline
\end{tabular}\end{center}
\end{table}

\section{Conclusions}

In this paper, we will develop a Frank-Wolfe type algorithm for the A-optimal experimental design problem. Our approach is similar to the Frank-Wolfe type algorithms developed for the D-optimal experimental design problem. Nevertheless, we are the first to discuss global and local convergence of the algorithms rigorously for the A-optimal experimental design problems.

\section{Acknowledgements}
The author would like to express her gratitude to Prof. Mike Todd for bringing the experimental design problem to her attention during her PhD candidacy at Cornell University and his constant support and mentorship. She is also thankful to the two anonymous reviewers for their valuable comments and suggestions.

\end{document}